\documentclass[conference]{IEEEtran}
\ifCLASSINFOpdf
  \usepackage[pdftex]{graphicx}
  \graphicspath{{../}}
  \DeclareGraphicsExtensions{.pdf,.jpeg,.png}
\else
\fi
\usepackage{amsmath}
\usepackage{array}

\usepackage{multirow}
\usepackage{booktabs}
\begin{document}



\title{Probabilistic Available Delivery Capability Assessment of General Distribution Network with Renewables}


\author{\IEEEauthorblockN{Hao Sheng, \textit{Member, IEEE}}
\IEEEauthorblockA{School of Electrical and Computer Engineering\\
Cornell University\\
Ithaca, NY, 14850 USA\\
Email: hs646@cornell.edu}
\and
\IEEEauthorblockN{Xiaozhe Wang, \textit{Member, IEEE}}
\IEEEauthorblockA{Department of Electrical and Computer Engineering\\
McGill University\\
Montreal, Quebec, Canada\\
Email: xiaozhe.wang2@mcgill.ca}}


%


\maketitle


\begin{abstract}
Rapid increase of renewable energy sources and electric vehicles in utility distribution feeders introduces more and more uncertainties. To investigate how such uncertainties may affect the available delivery capability (ADC) of the distribution network, it is imperative to employ a probabilistic analysis framework. In this paper, a formulation for probabilistic ADC incorporating renewable generators and load variations is proposed; a computationally efficient method to solve the probabilistic ADC is presented, which combines the up-to-date sparse polynomial chaos expansion (PCE) and the continuation method. A numerical example in the IEEE 13 node test feeder is given to demonstrate the accuracy and efficiency of the proposed method.
\end{abstract}


%
\IEEEpeerreviewmaketitle


\section{Introduction}

The increasing integration of renewables (e.g., wind power and solar photovoltaic (PV)) and new forms of energy consumption (e.g., electric vehicles (EVs)) in utility distribution feeders introduces more and more uncertainties apart from conventional load variations. The resulting randomness in distribution system may affect the delivering capability of a distribution system. The concept of available delivery capability (ADC) proposed in \cite{Sheng14a} is used to describe the maximum power that can be delivered over the existing amount for which no thermal overloads, voltage violations, and voltage collapse occur. Nevertheless, the proposed ADC and the developed numerical method to solve it \cite{Sheng14a} did not account for the uncertainties brought by renewables or EVs, which in turn are essential and stringent considering the constantly increasing integration of these components. In this paper, we propose a probabilistic framework to investigate the impact of uncertainties on the ADC of distribution networks.

The probabilistic analysis method, particularly, the probabilistic power flow was first proposed in the 1970s \cite{Borkowska74}, \cite{Allan74}, and many efforts were made thereafter. Generally, these methods can be divided into two categories: simulation methods and analytical methods. Monte Carlo simulation (MCS) \cite{Yu09} is the most widely used in the first category due to its simplicity, yet its high computational effort may prohibit its applications. Attempting to release the computational burden, analytical methods are developed which utilize mathematical approximations and assumptions. Some of the representative analytical methods include the cumulant method \cite{Zhang04} and its extended versions \cite{Fan12} as well as point estimation method \cite{Su05}, \cite{Morales07}. Nevertheless, the probability distributions of the power flow responses cannot be directly acquired from either cumulant methods or point estimation methods \cite{Ren16}. 

Polynomial chaos expansion (PCE) is another popular analytical method, which represents the probabilistic response as a sum of orthogonal polynomial basis functions. It was originally formulated with Gaussian random variables and Hermite polynomials by Wiener \cite{Wiener38}. Xiu \cite{Xiu02} further generalized the method by using non-Gaussian random variables together with optimal basis functions under the Askey polynomial scheme. The popularity of the PCE method is mainly attributed to the following features: (i) It can be easily integrated with existing deterministic analysis tools in a "non-intrusive" way. (ii) Accurate estimations for the probability distribution and the associated statistics can be obtained with low computational cost. This method has been applied to the probabilistic power flow \cite{Ren16} and load margin problems \cite{Haesen09}, yet it suffers from 'the curse of dimensionality' as the number of random inputs increases. This issue can be alleviated by the sparse-adaptive scheme \cite{Blatman09} which employs the step-wise regression \cite{Weisberg80} or least angle regression (LARS) \cite{Efron04} to detect a subset of significant coefficients, and thus reduces the computational efforts. In the context of power system, the sparse PCE has been applied to investigate probabilistic power flow on distribution networks in \cite{Ni17}. In this paper, we combine the sparse PCE and the continuation method to assess the probabilistic ADC. The main contributions of this paper are as follows: 
\begin{itemize}
\item A comprehensive formulation of probabilistic ADC problems is developed in distribution network under various uncertainties from large integration of wind power, solar PV, and load variations.
\item A computationally efficient yet accurate methodology is proposed to evaluate the probabilistic ADC, which combines the up-to-date sparse polynomial chaos expansion and continuation method.
\item Accurate probabilistic characteristics of the ADC can be achieved by the proposed method with much less computational efforts. The obtained ADC readily demonstrates how the uncertainty affect the capability of the distribution network, and how much more renewable penetration or load increase the system can withstand with low risk.
\end{itemize}

The rest of the paper is organized as the following. Section II introduces the probabilistic models involving wind power, solar PV and loads. Section III presents the mathematical formulation of probabilistic ADC. Section IV describes the sparse PCE method and its implementation in assessing ADC. The detailed algorithm to assess the probabilistic ADC is presented in Section V. The simulation results on the modified IEEE 13 node test feeder is given in Section VI. Conclusions and perspectives are given in Section VII.



\section{Power System Probabilistic Modeling}
The probabilistic model of power injections may depend on the time range of the problem under study and the characteristic of physical components. For instance, for long time scales, the Weibull distribution \cite{Karki06} is a good fit to the observed wind speed empirical distribution in many locations around the world, yet it is not considered as a good fit in time scales shorter than 10 min \cite{Zarate-Minano13}. In this paper, we choose the probabilistic distributions that are appropriate for short-term operational analysis.

\subsection{Wind Generation}
Suppose the forecasted value of wind speed is available from a wind speed forecaster, the probabilistic model of wind speed for short-term operational analysis can be represented as a normal distribution \cite{Ran16}
\begin{equation}
\label{eq:wind_pdf_normal}
f(v)=\frac{1}{\sqrt{2\pi}{{\sigma}_{v}}} \exp \left( -\frac{{{(v-{\mu}_{v})}^{2}}}{2{\sigma}_{v}^{2}} \right)
\end{equation}
where the mean ${{\mu}_{v}}$ is the forecasted wind speed, and variance ${{\sigma}_{v}}$ represents the forecasting error which can be obtained from historical data.

The wind speed-power output relation is defined as \cite{Roy02}
\begin{equation}
\label{eq:wind_p_val}
{{P}_{w}}(v)=\left\{ \begin{array}{*{35}{l}}
0 & v\le {{v}_{in}}  \\
\displaystyle \frac{v-{{v}_{in}}}{{{v}_{rated}}-{{v}_{in}}}{{P}_{r}} & {{v}_{in}}<v\le {{v}_{rated}}  \\
{{P}_{r}} & {{v}_{rated}}<v\le {{v}_{out}} \\
0 & v>{{v}_{out}}  \\
\end{array} \right.
\end{equation}
where ${{v}_{in}}$, ${{v}_{out}}$ and ${{v}_{rated}}$ are the cut-in, cut-out, and rated wind speed ($m/s$), ${{P}_{r}}$ is the rated wind power ($kW$) and ${{P}_{w}}$ is the active power output of wind power. The reactive power output ${{Q}_{w}}$ can be computed under the assumption of constant power factor (e.g., 0.85 lagging).

\subsection{Solar Generation}
For long time analysis, the solar radiation is typically presented by Beta distribution \cite{Salameh95}, while for short-term operational analysis with forecasted value, it can be described by normal distribution \cite{Ran16}
\begin{equation}
\label{eq:solar_pdf_normal}
f(r)=\frac{1}{\sqrt{2\pi}{{\sigma}_{r}}} \exp \left( -\frac{{(r-{{\mu}_{r}})}^{2}}{2{\sigma}_{r}^{2}} \right)
\end{equation}
where ${{\mu}_{r}}$ is the mean of solar radiation, and ${{\sigma}_{r}}$ is the standard deviation describing the forecasting error. The solar radiation-power output relation is defined as \cite{Park09}, \cite{Marwali98}
\begin{equation}
\label{eq:solar_p_val}
{{P}_{pv}}(r)=\left\{ \begin{array}{*{35}{l}}
\displaystyle \frac{{{r}^{2}}}{{{r}_{c}}{{r}_{std}}}{{P}_{r}} & 0\le r<{{r}_{c}}  \\
\displaystyle \frac{r}{{{r}_{std}}}{{P}_{r}} & {{r}_{c}}<r\le {{r}_{std}}  \\
{{P}_{r}} & r>{{r}_{std}}  \\
\end{array} \right.
\end{equation}
where $r$ is the forecasted solar radiation in $W/m^2$, ${{r}_{c}}$ is a certain radiation point set usually as 150 $W/m^2$,  ${{r}_{std}}$ is the solar radiation in the standard environment, ${{P}_{r}}$ is the rated capacity of the solar PV. Solar generation is injected into the grid at unity power factor \cite{WECC10}, and hence ${{Q}_{pv}}$ is assumed to be zero in this study.

\subsection{Load Variation}
The uncertainty of load power is usually assumed to follow the normal distribution \cite{Billinton08}, the probability density function of load active power is expressed as:
\begin{equation}
\label{eq:load_pdf}
f({{P}_{L}})=\frac{1}{\sqrt{2\pi}{{\sigma}_{P}}} \exp \left( -\frac{{{({{P}_{L}}-{{\mu}_{P}})}^{2}}}{2\sigma_{P}^{2}} \right)
\end{equation}
where the forecasted mean value ${{\mu}_{P}}$ of ${{P}_{L}}$ is provided by load forecaster, and ${{\sigma}_{P}}$ denotes the forecasting error. Generally, the load forecaster only provides the active power, whereas the reactive power is determined under the assumption of constant power factor.


\section{Mathematical Formulation of Probabilistic ADC}
The three-phase distribution power flow equations can be represented as
\begin{equation}
\label{eq:power_flow}
f\left(x\right)=\begin{bmatrix}
    {P_{i0}^{\varphi}-P_{i}^{\varphi}(x)} \\ 
    {Q_{i0}^{\varphi}-Q_{i}^{\varphi}(x)}
\end{bmatrix}=0
\end{equation}
where $x={{\left[ {{\theta}_{a}},{{\theta}_{b}},{{\theta}_{c}},{{V}_{a}},{{V}_{b}},{{V}_{c}} \right]}^{T}}$, e.g., voltage angles and magnitudes for all phases.

We can simulate the quasi-static behaviors of distribution systems under various uncertainties. Given a forecasted wind speed vector $v$, a solar radiation vector $r$, and a forecasted load vector ${{P}_{L}}$, the three phase probabilistic continuation power flow (PCPF) equations of a $N$ bus system can be explicitly expressed as follows. For PQ type nodes, the PCPF equations are:
\setlength{\arraycolsep}{0.0em}
\begin{eqnarray}
\label{eq:pq_bus_p}
P_{i0}^{\varphi} & - & V_{i}^{\varphi}\sum\limits_{j=1}^{N}{\sum\limits_{k=1}^{M}{V_{j}^{k}\left(G_{ij}^{\varphi k}\cos \theta_{ij}^{\varphi k}+B_{ij}^{\varphi k}\sin \theta_{ij}^{\varphi k} \right)}} \\
& + & \lambda \left(\Delta P_{Gi}^{\varphi}+\Delta P_{wi}^{\varphi}({{v}_{i}})+\Delta P_{pvi}^{\varphi}({{r}_{i}})-\Delta P_{Li}^{\varphi}(P_{Li}^{\varphi})\right)=0 \nonumber
\end{eqnarray}
\setlength{\arraycolsep}{5pt}
\setlength{\arraycolsep}{0.0em}
\begin{eqnarray}
\label{eq:pq_bus_q}
Q_{i0}^{\varphi} & - & V_{i}^{\varphi}\sum\limits_{j=1}^{N}{\sum\limits_{k=1}^{M}{V_{j}^{k}\left(G_{ij}^{\varphi k}\sin \theta_{ij}^{\varphi k}-B_{ij}^{\varphi k}\cos \theta_{ij}^{\varphi k} \right)}} \nonumber \\ 
& + & \lambda \left(\Delta Q_{Gi}^{\varphi}+\Delta Q_{wi}^{\varphi}({{v}_{i}})-\Delta Q_{Li}^{\varphi}(P_{Li}^{\varphi}) \right)=0
\end{eqnarray}
\setlength{\arraycolsep}{5pt}
For PV type nodes, the corresponding PCPF equations are:
\setlength{\arraycolsep}{0.0em}
\begin{eqnarray}
\label{eq:pv_bus_p}
P_{i0}^{\varphi} & - & V_{i}^{\varphi}\sum\limits_{j=1}^{N}{\sum\limits_{k=1}^{M}{V_{j}^{k}\left(G_{ij}^{\varphi k}\cos \theta _{ij}^{\varphi k}+B_{ij}^{\varphi k}\sin \theta _{ij}^{\varphi k} \right)}} \\
& + & \lambda \left(\Delta P_{Gi}^{\varphi}+\Delta P_{wi}^{\varphi}({{v}_{i}})+\Delta P_{pvi}^{\varphi}({{r}_{i}})-\Delta P_{Li}^{\varphi}(P_{Li}^{\varphi})\right)=0 \nonumber
\end{eqnarray}
\setlength{\arraycolsep}{5pt}
\begin{equation}
\label{eq:pv_bus_v}
V_{i}^{\varphi}={{V}_{i0}}
\end{equation}
\setlength{\arraycolsep}{0.0em}
\begin{eqnarray}
\label{eq:pv_bus_q}
Q_{Gi}^{\varphi} & - & Q_{Li0}^{\varphi}-V_{i}^{\varphi}\sum\limits_{j=1}^{N}{\sum\limits_{k=1}^{M}{V_{j}^{k}\left(G_{ij}^{\varphi k}\sin \theta_{ij}^{\varphi k}-B_{ij}^{\varphi k}\cos \theta_{ij}^{\varphi k} \right)}} \nonumber \\ 
& + & Q_{wi0}^{\varphi} + \lambda \left(\Delta Q_{wi}^{\varphi}({{v}_{i}})-\Delta Q_{Li}^{\varphi}(P_{Li}^{\varphi}) \right)=0
\end{eqnarray}
\setlength{\arraycolsep}{5pt}
\begin{equation}
\label{eq:pv_bus_qlimit}
{{Q}_{min,i}}\le Q_{Gi}^{\varphi}\le {{Q}_{max,i}}
\end{equation}
where $G_{ij}^{\varphi k}$ and $B_{ij}^{\varphi k}$ are entries in the bus admittance matrix; $\Delta P_{wi}^{\varphi}({{v}_{i}})$, $\Delta P_{pvi}^{\varphi}({{r}_{i}})$, $\Delta P_{Li}^{\varphi}$ and $\Delta P_{Gi}^{\varphi}$ are the real power variation from wind power, solar PV, load and other types of DG at the phase $\varphi$ of bus $i$ respectively; $\Delta Q_{wi}^{\varphi}({{v}_{i}})$ and $\Delta Q_{Li}^{\varphi}$ are the reactive power variation from wind power and load, respectively; $Q_{Gi}^{\varphi }$ is the reactive generation, and $M$ is the number of phases. If $Q_{Gi}^{\varphi}$ exceeds its limits, say ${{Q}_{min,i}}$ or ${{Q}_{max,i}}$, then the terminal bus switches from PV to PQ with $Q_{Gi}^{\varphi}$fixed at the violated limit.

In fact, the set of parameterized three-phase PCPF equations (\ref{eq:pq_bus_p})-(\ref{eq:pv_bus_qlimit}) can be described in the following compact form
\begin{equation}
\label{eq:cpf_equation}
f\left( x,\mu,\lambda,u \right)=f\left( x,\mu \right)-\lambda b(u)=0
\end{equation}
where $x$ is the state vector, $\mu$ is the control parameters vector such as the tap ratio of transformer, $u=\left[ v,r,{{P}_{L}} \right]$ is the random vector describing the wind speed, the solar radiation, and the load active power. Besides, the load-generation variation vector $b$ of the system is
\begin{equation}
\label{eq:cpf_variation}
b\left( u \right)=\begin{bmatrix}
   \Delta P_{G}^{\varphi}+\Delta P_{w}^{\varphi}(v)+\Delta P_{pv}^{\varphi}(r)-\Delta P_{L}^{\varphi}({{P}_{L}})  \\
   \Delta Q_{G}^{\varphi}+\Delta Q_{w}^{\varphi}(v)-\Delta Q_{L}^{\varphi}({{P}_{L}})  \\
\end{bmatrix}
\end{equation}

Assuming the power factors are constant, the reactive power variation $\Delta Q_{wi}^{\varphi}$ and $\Delta Q_{Li}^{\varphi}$ can be calculated by
\begin{equation}
\label{eq:wind_q_val}
\Delta Q_{wi}^{\varphi}({{v}_{i}})=\tan {{\theta}_{\alpha i}}\Delta P_{wi}^{\varphi}({{v}_{i}})
\end{equation}
\begin{equation}
\label{eq:load_q_val}
\Delta Q_{Li}^{\varphi}({{P}_{L}})=\tan {{\theta}_{\beta i}}\Delta P_{Li}^{\varphi}({{P}_{L}})
\end{equation}
where ${{\theta}_{\alpha i}}$ and ${{\theta}_{\beta i}}$ are the power factors of wind power and load, respectively. It is obvious that the set of the parameterized power flow equations become the base-case power flow equation if $\lambda =0$.

The probabilistic ADC formulation therefore can be proposed as the following:
\begin{equation}
\label{eq:prob_adc}
\begin{aligned}
& \text{max}
& & \lambda \\
& \text{s.t.} & & f\left( x,\mu \right)-\lambda b(u)=0 & (a) \\
& & & {{V}_{min}}\le V_{i}^{\varphi}\left( x,\mu,\lambda,u \right)\le {{V}_{max}} & (b) \\
& & & I_{ij}^{\varphi}\left( x,\mu,\lambda,u \right)\le {{I}_{ij,max}} & (c) \\
& & & {{Q}_{min,i}}\le Q_{Gi}^{\varphi }\left( x,\mu,\lambda,u \right)\le {{Q}_{max,i}} & (d)
\end{aligned}
\end{equation}
where ${{V}_{min}}$ and ${{V}_{max}}$ are the lower and upper limits of bus voltages; ${{I}_{ij,max}}$ is the specified capacity of the line or transformer between bus $i$ and bus $j$. $\lambda$ is the normalized load margin under given load-generation variation vector. The maximum value of λ that could be achieved without the violation of (\ref{eq:prob_adc}) corresponds to the ADC.  Note that $\lambda$ is a random variable due to the random input $u$. Equation (a) specifies that the solution must satisfy the parameterized power flow equations (\ref{eq:cpf_equation}); Equations (b)-(d) imply that the solution has to satisfy typical operational and electrical constrains.


\section{Probabilistic ADC Assessment Using Sparse Polynomial Chaos Expansion}
The generalized PCE method \cite{Xiu02} may use different polynomial chaos basis depending on the specific probability distribution of the random variables. In this paper, the Hermite polynomials are applied due to the fact that all random inputs in (\ref{eq:prob_adc}) follow the normal distribution. In this section, the proposed sparse PCE method for probabilistic ADC assessment is presented which integrates a sparse scheme into the original PCE method \cite{Isukapalli98} to handle a large number of random inputs.

\subsection{Transformation of Random Inputs}
Suppose $\xi =[{{\xi}_{1}},{{\xi}_{2}},...,{{\xi}_{n}}]$ is the set of standard normal random variables, each non-standard normal input ${{u}_{i}}$ can be expressed by the following quantile function
\begin{equation}
\label{eq:quant_func}
{{u}_{i}}=F_{i}^{-1}\left(\Psi({{\xi}_{i}}) \right)
\end{equation}
where $F_{i}^{-1}$ is the inverse cumulative probability function of ${{u}_{i}}$. $\Psi$ represents the cumulative distribution function of standard normal distribution.

\subsection{Functional Approximation of Desired Responses}
Given the set of $n$-dimensional standard normal random variables $\xi =[{{\xi}_{1}},{{\xi}_{2}},...,{{\xi}_{n}}]$, the desired responses $y$, say the ADC w.r.t. voltage violation, thermal violation, and voltage collapse in this paper, can be approximated by the expansion
\setlength{\arraycolsep}{0.0em}
\begin{eqnarray}
\label{eq:pce_full}
y & = & {{c}_{0}} + \sum\limits_{{{i}_{1}}=1}^{n}{{{c}_{{{i}_{1}}}}{{H}_{1}}\left( {{\xi}_{{{i}_{1}}}} \right)}+\sum\limits_{{{i}_{1}}=1}^{n}{\sum\limits_{{{i}_{2}}=1}^{{{i}_{1}}}{{{c}_{{{i}_{1}}{{i}_{2}}}}{{H}_{2}}\left( {{\xi}_{{{i}_{1}}}},{{\xi}_{{{i}_{2}}}} \right)}} \nonumber \\
& + & \sum\limits_{{{i}_{1}}=1}^{n}{\sum\limits_{{{i}_{2}}=1}^{{{i}_{1}}}{\sum\limits_{{{i}_{3}}=1}^{{{i}_{2}}}{{{c}_{{{i}_{1}}{{i}_{2}}{{i}_{3}}}}{{H}_{3}}\left( {{\xi}_{{{i}_{1}}}},{{\xi}_{{{i}_{2}}}},{{\xi}_{{{i}_{3}}}} \right)}}}+...
\end{eqnarray}
\setlength{\arraycolsep}{5pt}
with unknown coefficients ${{c}_{0}}$, ${{c}_{{{i}_{1}}}}$, ${{c}_{{{i}_{1}}{{i}_{2}}}}$ and ${{c}_{{{i}_{1}}{{i}_{2}}{{i}_{3}}...}}$. The multidimensional Hermite polynomials ${{H}_{p}}$ of order $p$ can be computed by the following formula \cite{Isukapalli98}
\begin{equation}
\label{eq:hermite}
{{H}_{p}}\left( {{\xi}_{{{i}_{1}}}},...,{{\xi}_{{{i}_{p}}}} \right)={{\left( -1 \right)}^{p}}{{e}^{0.5{{\xi}^{T}}\xi}}\frac{{{\partial}^{p}}}{\partial{{\xi}_{{{i}_{1}}}}...\partial{{\xi}_{{{i}_{p}}}}}{{e}^{-0.5{{\xi}^{T}}\xi}}
\end{equation}

Generally, higher accuracy can be achieved with a higher order, yet higher computational effort. In practice, the order $p$ should be properly chosen to achieve a tradeoff between the accuracy and the computational cost. The recommended value of $p$ is 2 or 3 \cite{Haesen09}. In this paper, the second order is applied, and the corresponding truncated expansion is
\begin{equation}
\label{eq:pce_trunc}
y={{c}_{0}}+\sum\limits_{i=1}^{n}{{{c}_{i}}{{\xi}_{i}}}+\sum\limits_{i=1}^{n}{{{c}_{ii}}\left( {\xi}_{i}^{2}-1 \right)}+\sum\limits_{i=1}^{n-1}{\sum\limits_{j>i}^{n}{{{c}_{ij}}{{\xi}_{i}}{{\xi}_{j}}}}
\end{equation} 

The number of unknown PCE coefficients $K$ is
\begin{equation}
\label{eq:pce_coeff}
K=\begin{pmatrix}
n+p  \\
p  \\
\end{pmatrix}=\frac{(n+p)}{n!p!}
\end{equation}
\subsection{Estimation of PCE Coefficients}
To estimate the coefficients in truncated expansion (\ref{eq:pce_trunc}), one need to run simulations on a set of elaborately selected sample points of $\xi$ to get the accurate response vector $y$. The efficient collocation method proposed in \cite{Isukapalli98} is applied in this work which turns to be effective. The key idea is that the collocation points are selected in a way that they can capture high probability regions to ensure accuracy. To this end, the union of zero and the root of $p+1$ order Hermite polynomial are selected for the Hermite polynomial of $p$. For each sample of $\xi$, the corresponding random inputs $u$ will be computed by (\ref{eq:quant_func}) and taken into (\ref{eq:prob_adc}) to compute the accurate responses $y$ using CDFLOW \cite{Sheng14b}. After deterministic simulation on all sample points has completed, the least square estimate of the coefficients can be achieved by
\begin{equation}
\label{eq:least_sqr}
{{A}^{T}}Ac={{A}^{T}}y
\end{equation}
where $A$ is called the design matrix, each row corresponds to a sample of $\xi$, $c$ is the coefficient vector. To solve (\ref{eq:least_sqr}) directly, one needs to make sure that $A$ has full column rank. Note that the number of simulations to solve the coefficient is typically small yet depends on the number of random inputs.

The limitation of the truncated expansion in (\ref{eq:pce_trunc}) is that for a problem with a large number of random inputs, the number of unknown coefficients $K$ is fairly large, and the consequent computational cost of running simulations for solving the coefficients may become unaffordable. To mitigate this issue, we integrate a sparse scheme using the least angle regression (LARS) into (\ref{eq:pce_trunc}). The key idea of LARS is to automatically detect the significant coefficients so that they can be estimated by a small set of simulations, while the rest of coefficients are set to zero. This scheme is based on the fact that the predictors (column vector of $A$ in (\ref{eq:least_sqr})) are not equally relevant in the sense that some predictors may contribute more significantly to the response $y$ than the others. LARS is an effective regression tool for fitting the linear model even when the number of predictors is much larger than the available simulation data (rows of $A$). The readers are referred to \cite{Efron04} for more details on LARS.

\subsection{Statistical Analysis}
After the coefficients are determined, the sample space can be extensively sampled, and the probabilistic ADC can be easily evaluated by the polynomial approximant (\ref{eq:pce_trunc}) without solving (\ref{eq:prob_adc}) which is computationally expensive. Then the cumulative distribution function (CDF) and typical statistics such as mean, variance, skewness, kurtosis, and confidence interval can be readily computed.


\section{Computation of Probabilistic ADC}
In this section, a step-by-step description of the proposed probabilistic ADC calculation is summarized below:

Step 1: Input network data and forecasted data including wind speed, solar radiation, and load, as well as their standard deviation from historical data, build up the load-generation variation vector $b(u)$.

Step 2: Apply the efficient collocation method to select ${{M}_{C}}$ samples of the standard random variables $\xi$, and then transform them into samples of the inputs $u$ by (\ref{eq:quant_func}). 

Step 3: Run deterministic CDFLOW on the samples of $u$ to get the accurate response $y$, i.e., solve (\ref{eq:prob_adc}) for ADC w.r.t. voltage violation, thermal violation, and voltage collapse.

Step 4: For each of the response in $y$, apply LARS to select out ${{M}_{C}}$ columns from the $K$ columns of the design matrix $A$, then estimated the corresponding ${{M}_{C}}$ coefficients by the least square method, the rest of coefficients are set to 0.

Step 5: Once the coefficients of PCE are solved, sample $\xi$ extensively, e.g., ${{M}_{S}}$ samples, and apply the solved functional approximant (\ref{eq:pce_trunc}) to evaluate the corresponding responses $y$ for all these samples.

Step 6. Compute the statistics of each response, and generate the result report.

Remark: the number of samples ${{M}_{C}}$ in Step 3 is usually much smaller than ${{M}_{S}}$ in Step 5. Unlike MCS, PCE does not solve (\ref{eq:prob_adc}) for all ${{M}_{S}}$ samples in Step 5, hence it is more efficient. The main computational effort of PCE lies in Step 3.


\section{Numerical Studies}
In this section, we apply the proposed method to investigate the probabilistic ADC of the modified IEEE 13 node test feeder \cite{IEEE92}. The Monte Carlos simulation is used as a benchmark to validate the accuracy and the performance of the proposed method. We add two solar PVs and two wind generators to the IEEE 13 node test feeder, the total loads of which are 1.733 MW, 1.051 Mvar. The forecasted wind speed and parameters of wind generators are shown in Table \ref{tab:wind_para}, and the forecasted solar radiation and parameters of the Solar PVs are listed in Table \ref{tab:solar_para}. The stochastic variations of 8 single phase loads at 6 load buses are set to be the base case value with a standard deviation of 5\% of their mean values.
\begin{table}[]
\renewcommand{\arraystretch}{1.3}
\caption{Parameters of the Wind Generators}
\label{tab:wind_para}
\centering
\begin{tabular}{|c|c|c|c|c|c|c|c|c|}
    \hline
    \bfseries Bus & \bfseries Phase & $v$ & $\mu_{v}$ & $P_{r}$ & $v_{rate}$ & $v_{in}$ & $v_{out}$ & $v_{0}$ \\
    \hline
    680 & ABC & 10.0 & 0.6 & 450.0 & 15.0 & 4.0 & 25.0 & 6.0 \\
    \hline
    634 & ABC & 10.0 & 0.6 & 300.0 & 15.0 & 4.0 & 25.0 & 6.0 \\
    \hline
\end{tabular}
\end{table}
\begin{table}[]
\renewcommand{\arraystretch}{1.3}
\caption{Parameters of the solar PVs}
\label{tab:solar_para}
\centering
\begin{tabular}{|c|c|c|c|c|c|c|c|}
    \hline
    \bfseries Bus & \bfseries Phase & $r$ & $\mu_{r}$ & $P_{r}$ & $r_{c}$ & $r_{std}$ & $r_{0}$ \\
    \hline
    675 & ABC & 500.0 & 25.0 & 180 & 150 & 1000 & 300.0 \\
    \hline
    692 & ABC & 500.0 & 25.0 & 240 & 150 & 1000 & 300.0 \\
    \hline
\end{tabular}
\end{table}

\begin{figure}
\centering
\includegraphics[width=2.5in]{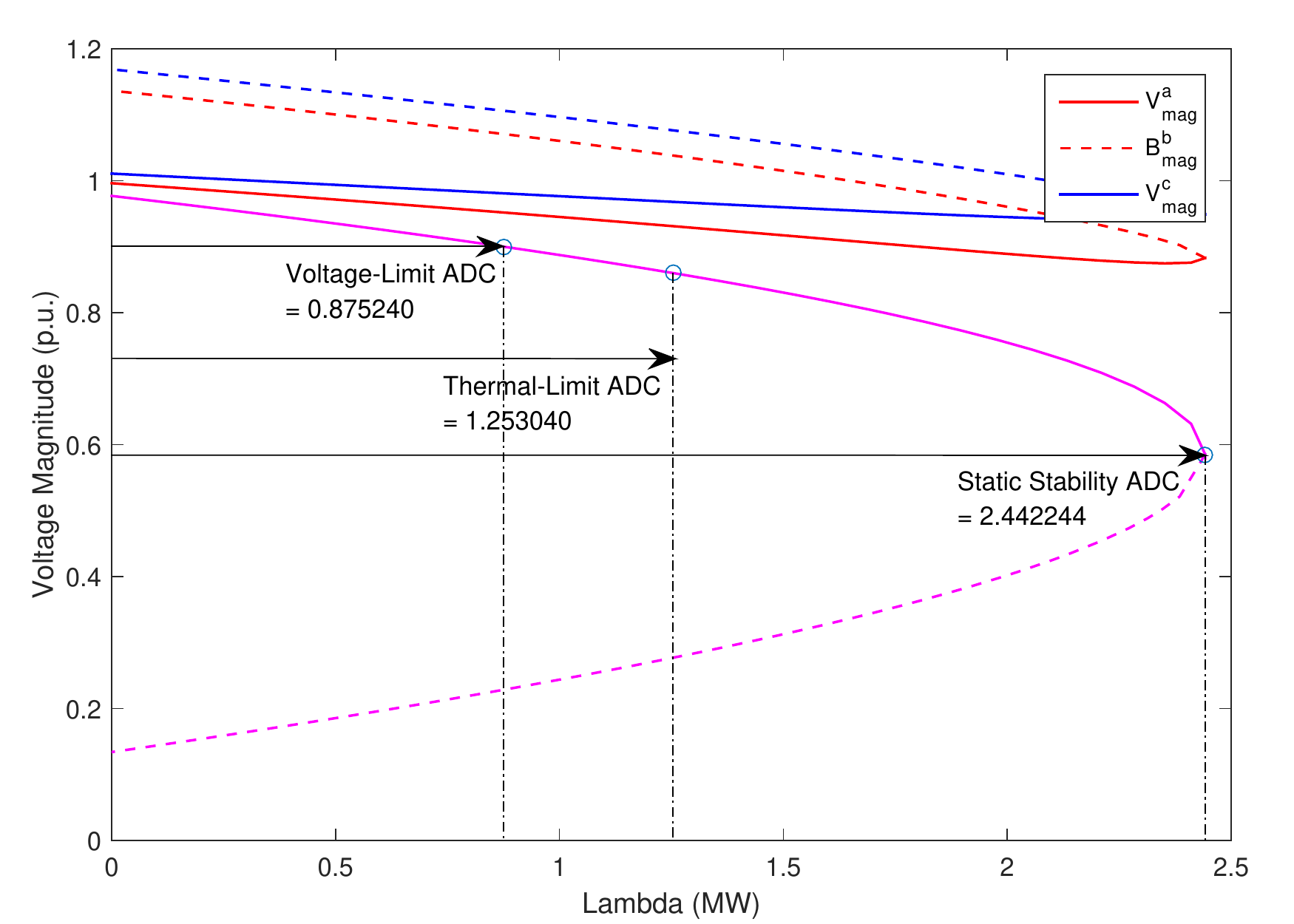}
\caption{The voltage-limit ADC is 0.875 MW, the thermal-limit ADC is 1.253 MW and the static-stability-limit ADC is 2.442 MW. The (overall) ADC is hence 0.875 MW.}
\label{fig:adc_base}
\end{figure}

Figure \ref{fig:adc_base} shows a P-V curve of the three-phase bus 675 at the sample point where all random inputs are equal to their mean values. Among the three ADCs, the ADC subject to voltage violation, 0.875 MW, is the smallest one due to fact that the voltage at bus 611 reaches the lower limit 0.90 p.u. Hence the overall ADC of is 0.875 MW.

\begin{figure}
\centering
\includegraphics[width=2.5in]{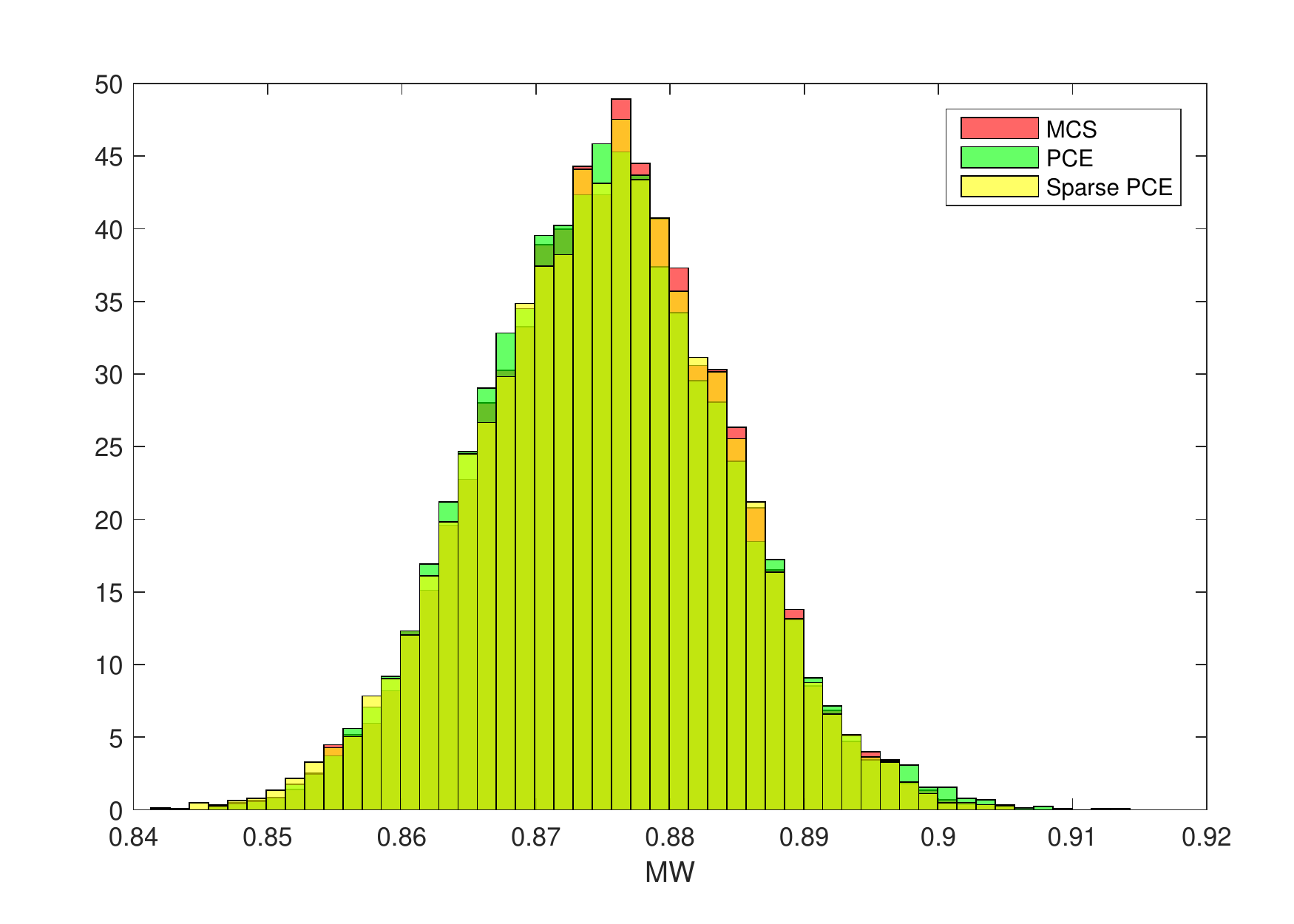}
\caption{The distribution of ADC subject to voltage violation computed by MCS, PCE, and sparse PCE. They are close to each other.}
\label{fig:adc_dist}
\end{figure}

Next, we assess the probabilistic ADC by the proposed method and compare the results with the benchmark MCS. After the coefficients of PC expansion (\ref{eq:pce_trunc}) are solved, 10000 samples are generated to assess the probabilistic ADC by the PCE approximant, the sparse PCE (SPCE) approximant, respectively. In this case, the overall ADC corresponds to the one subject to voltage violation. Figure \ref{fig:adc_dist} shows the estimated probabilistic distribution of the ADC by the MCS, the PCE and the sparse PCE, from which we can see that both PCE and sparse PCE are able to provide reasonably good estimations. Similar results are achieved when comparing the estimated statistics of the ADC by the three methods as shown in Table \ref{tab:adc_comp}. Nevertheless, to get these comparable accuracy, MCS needs to run 10000 simulations (i.e. solving (\ref{eq:prob_adc})), while the PCE needs 91 simulations, and the sparse PCE only requires 31 simulations.
\begin{table}[]
\renewcommand{\arraystretch}{1.3}
\caption{Comparison of the Estimated Statistics of the ADC}
\label{tab:adc_comp}
\centering
\begin{tabular}{|c|c|c|c|c|c|}
\hline
\multirow{2}{*}{\bfseries ADC} & \multirow{2}{*}{\bfseries Method} & \multicolumn{4}{c|}{\bfseries Statistics of Probabilistic ADC} \\
\cline{3-6}
                                     &      & $Mean$   & $Var.$ & $Skew.$ & $Kurt.$ \\
\hline
\multirow{3}{1cm}{Voltage violation} & MCS  & 0.875229 & 0.000080 & -0.013951 & 2.975912 \\
\cline{2-6}
                                     & PCE  & 0.875029 & 0.000084 &  0.127900 & 3.079851 \\
\cline{2-6}
                                     & SPCE & 0.875016 & 0.000083 & -0.067090 & 2.995516 \\
\hline
\multirow{3}{1cm}{Thermal violation} & MCS  & 1.251807 & 0.000423 & -0.042778 & 3.005286 \\
\cline{2-6}
                                     & PCE  & 1.252264 & 0.000434 &  0.012322 & 3.024292 \\
\cline{2-6}
                                     & SPCE & 1.252278 & 0.000432 & -0.048233 & 3.009263 \\
\hline
\multirow{3}{1cm}{Voltage collapse}  & MCS  & 2.442311 & 0.000517 &  0.005444 & 2.974035 \\
\cline{2-6}
                                     & PCE  & 2.442316 & 0.000517 &  0.002986 & 2.972921 \\
\cline{2-6}
                                     & SPCE & 2.442368 & 0.000517 & -0.008261 & 2.968473 \\
\hline
\end{tabular}
\end{table}

\begin{figure}
\centering
\includegraphics[width=2.5in]{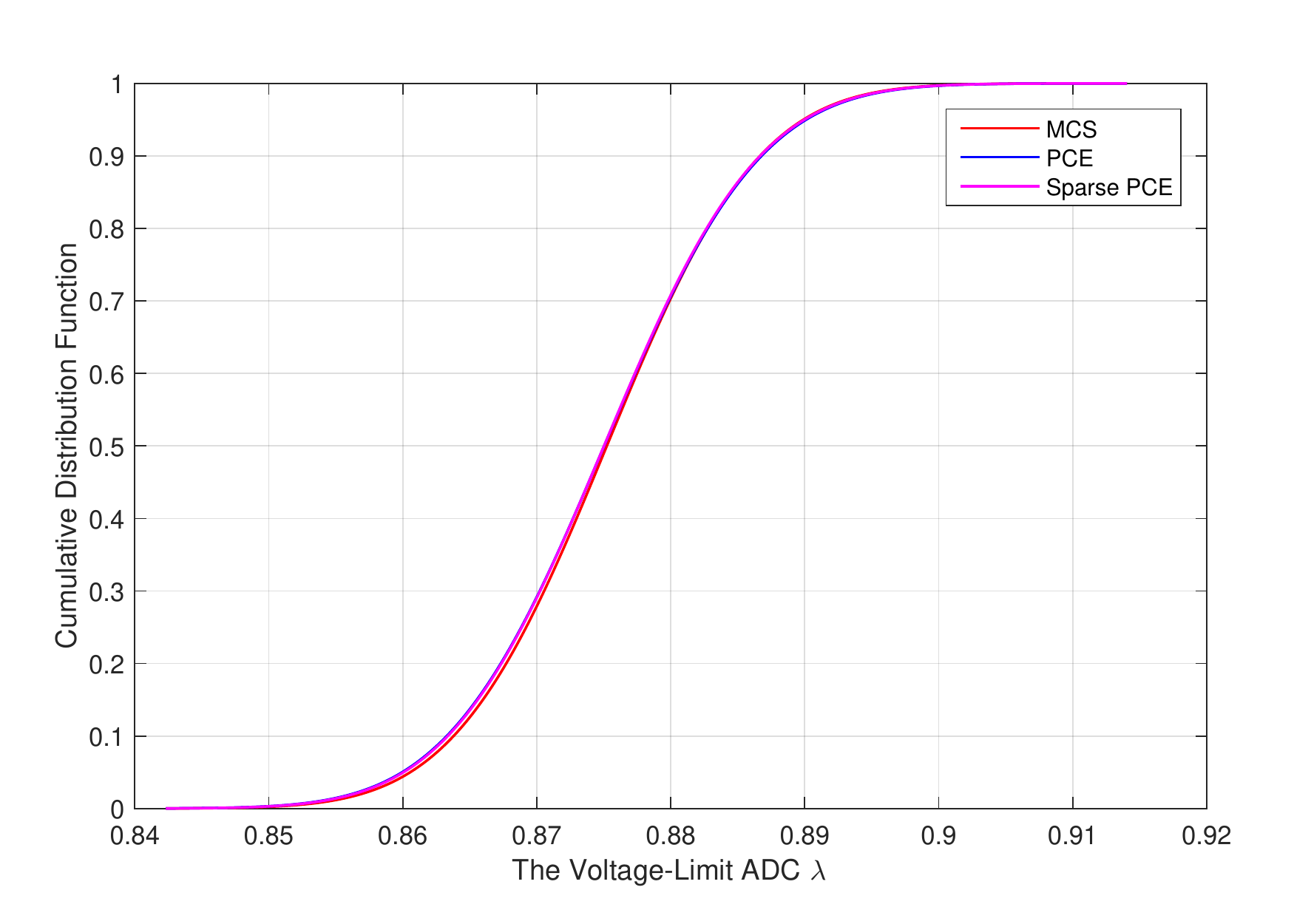}
\caption{The cumulative distribution curve of ADC w.r.t. voltage violation computed by MCS, PCE, and sparse PCE. They are almost overlapped.}
\label{fig:adc_cdf}
\end{figure}

In addition, the cumulative distribution curve of the ADC subject to voltage violation computed by MCS, PCE, and sparse PCE are shown in Figure \ref{fig:adc_cdf}. They are almost overlapped indicating that the proposed sparse PCE method possesses a good accuracy in estimating the cumulative distribution curve. Similar results for ADC subject to thermal violation and voltage collapse can be obtained.

However, it’s worth mentioning that the number of deterministic simulations required by the PCE method (i.e., ${{M}_{C}}$ in Step 2 in Section V) grows dramatically as the number of random inputs increases. This issue has been mitigated by integrating LARS to reduce the number of simulations, but the computational cost for LARS itself grows quickly for large problems because many matrix operations are involved. Besides, the computational effort for solving the coefficients is proportional to the number of desired responses because each response has an independent set of coefficients to be solved. Hence, PCE may not be superior to MCS in large systems. Further investigations on computational efforts are needed and are our future work plan.

\section{Conclusion and Perspectives}
In this paper, we have proposed a formulation of probabilistic ADC to incorporate various uncertainties introduced by wind power, solar PV and loads. A computationally efficient method to compute the probabilistic ADC is also developed, which combines the up-to-date sparse PCE and the continuation method. The proposed sparse PCE method is able to accurately compute the probabilistic characteristics of the ADC with much less computational effort compared with the MCS.

The probabilistic ADC provides a comprehensive information and intuitive picture regarding how the uncertainties affect the delivery capabilities of the distribution network, and how much more renewable penetration or load increase the system can withstand with low risk. In the future, we plan to develop control measures to reduce the variance of ADC and increase ADC by mitigating the violations at weak buses and branches. We believe that designing control actions to reduce the variance with limited resources will be a fruitful future development.









\bibliographystyle{IEEEtran}
%

\end{document}